\documentstyle[12pt]{article}

\catcode`\@=11

\@addtoreset{equation}{section}

\begin{document}
          
\baselineskip 24pt
\newcommand{\numero}{SWAT/33}
%Enter SHEP preprint number
\newcommand{\titre}{FERMION MASS POSTDICTIONS IN A GENERALIZED}
\newcommand{\titreb}{EXTENDED TECHNICOLOUR SCENARIO}
\newcommand{\auteura}{Nick Evans}
\newcommand{\addressa}{ }
\newcommand{\auteurc}{D.A. Ross }
\newcommand{\beq}{\begin{equation}}
\newcommand{\eeq}{\end{equation}}
\newcommand{\Fn}{\mbox{$F(p^2,\Sigma)$}}

\newcommand{\addressc}{Department Of Physics  \\   University   of
     Wales, Swansea\\ Singleton Park, Swansea \\ SA2 8PP, U.K. }
\newcommand{\abstrait}{We review the recent discussion in the literature of 
one family extended technicolour models with techni-fermion mass spectra
compatible with the experimental data for the precision parameters S,T,U,V,
W and X and ETC interactions compatible with the LEP measurements of the 
$Z\rightarrow b{\bar b}$ vertex. To investigate whether these scenarios are
consistent with the third family fermion masses we develop a generalized 
ETC model in which ETC interactions are represented by four Fermi interactions.
We discuss in detail the reliability of the gap equation approximation to
the non-perturbative dynamics. 
Two generic scenarios of couplings fit the precision data and third family masses;
one is an unpredicitive existence proof, the other, which generates the large 
top mass by direct top condensation, has a minimal number of interactions 
that break the global symmetry of the light fermions in the observed manner.
This latter scenario makes surprisingly good predictions of the charm,
strange and up quark masses.
     }
          
\begin{titlepage}
\hfill \numero
\vspace{.5in}
\begin{center}
{\large{\bf \titre }}
{\large{\bf \titreb}}
\bigskip \\by\bigskip\\ \auteura \bigskip \\ \addressc \\

\renewcommand{\thefootnote}{ }
\vspace{.5 in}
{\bf Abstract}
\end{center}
\abstrait
\end{titlepage}

\def\id{\rlap{1}\hspace{0.15em}1}

\section{Introduction}

The Holy Grail of the next generation of accelerator experiments is a 
renormalizable, predictive model of the gauge boson and fermion masses that
break electroweak symmetry. Models
in which electroweak symmetry is broken by a condensate of strongly interacting
fermions \cite{TC,TMSM} are an enticing possibility since they appeal to 
the successes of the BCS theory of superconductivity and chiral
symmetry breaking in QCD. Whilst strongly interacting models 
such as technicolour \cite{TC} provide a simple explanation of electroweak
symmetry breaking (EWSB) and the W and Z gauge boson masses, the diverse light 
fermion masses are much harder to understand. In the past theorists have
tended to concentrate on building models that extended the basic technicolour
scenario \cite{ETC,ETC2,ETC3} to include the light fermion masses (extended 
technicolour models, ETC) as an existence proof that technicolour models can
generate the diverse spectrum observed. Many of these models \cite{ETC2}, by virtue of
being existence proofs, have been very complicated having at least as 
many free parameters as there are elements in the light fermion mass matrices.
The hope is that experimental discoveries will shed light on a simpler
model along these lines which predicts some or all of the light fermion 
masses.

Recent precision tests of particle interactions below the Z mass from LEP
experiments \cite{Burgess3} and low energy atomic measurements \cite{atomic} 
have tightly constrained 
the parameters in the low energy effective theory of the electroweak symmetry
breaking sector. The effects of particles heavier than $M_Z/2$ (which are
integrated from the effective theory probed at LEP) on low energy observables
have been neatly summerized in terms of the parameters S,T,U,V,W and X 
\cite{Peskin,Burgess} as
well as the deviation from the tree 
level prediction for the process $Z\rightarrow b{\bar b}$ 
\cite{ZBB,ZBB2,ZBB3}. This new data 
has been used to rule out many of the ETC models constructed prior to LEP.
Recent work \cite{ZBB3,Terning,Maj,NPS} has concentrated on finding techni-fermion 
mass spectra and
extended technicolour interactions that are consistent with the new precision
data. The conclusion has been that ETC scenarios with light techni-fermions
and ETC interactions broken above 10 GeV still provide valid existence proofs
of ``realistic" strongly interacting models of EWSB.

In this paper we wish to investigate whether the recent precision data sheds
light on the form of a simple predicitive technicolour scenario. In Section 2 
we review the analysis of the precision data and the conclusion as to the
form a realistic techni-fermion mass spectrum must take. In Section 3 we introduce a generalized
form of ETC model with a minimal number of new ETC interactions which we
represent by four Fermi interactions. In order to simplify the initial
treatment we set the weak mixing angles to zero and concentrate
on the charged leptons and quarks. The CKM matrix elements and neutrino 
mass generation are extremely model dependent; we stress that we make this
approximation in order to generate generic statements about ETC.
We discuss how in
principle such a model could be predictive. We also introduce a 
model of the existence
proof type in order to show that ETC can be coerced to fit any fermion 
mass data given sufficient new parameters (this scenario is completely 
unpredictive but included for completeness). To perform a numerical search
of the parameter spaces of these models we must make some approximation to
the full non-perturbative strong dynamics. We shall use the familiar gap 
equation approximation \cite{gap}. In Section 4 we review the successes and failures of 
the gap equation with some numerical examples. In Section 5 we present two
general scenarios of ETC model with techni-fermion and third family mass
spectra compatible with all available experimental data. One of these 
scenarios is entirely unpredictive whilst the other, a simple ETC model 
with top condensation makes surprisingly good predictions for the up, charm and
strange quark masses. We present these predictions in Section 6. Finally
in Section 7 we conclude by discussing the implications of our generalized
model for ETC model building and the need to extend the analysis to the
neutrino sector and the CKM matrix elements.

\section{Precision Constraints On ETC}

Recent precision LEP data \cite{Burgess3} and low energy atomic physics measurements 
provide stringent constraints on the physics responsible for EWSB. In this
section we review these constraints and the results of Refs \cite{ZBB3,Terning,Maj,NPS} 
which suggest ETC models compatible with these constraints may exist. We
divide our discussion of these constraints into two types: oblique
corrections and  non-oblique corrections. In addition we briefly discuss constraints
from flavour changing neutral currents (FCNC) and a light pseudo-scalar spectrum.

\subsection{Oblique Corrections}

The major contributions to low energy observables from fermions and scalars
with masses greater than $M_Z/2$ occur at one loop as oblique corrections 
to gauge boson propagators \cite{Lynn}. These corrections have been 
parameterized by Peskin and Takeuchi \cite{Peskin} and by Burgess {\it et al.} 
\cite{Burgess} in terms of the
six parameters S,T,U,V,W and X.  LEP's precision measurements have been 
performed
on the Z mass resonance and hence the parameters associated with charged current 
interactions, U and W, are the least well constrained experimentally. 
A global fit \cite{Burgess2} to the experimental data in which all
six parameters S,T,U,V,W and X are allowed to vary simultaneously gives
the one standard deviation bounds

\beq \begin{array}{ccccccc}
S & \sim & -0.93 \pm 1.7 & \hspace{2cm} & V & \sim & 0.47 \pm 1.0 \\
&&\\
T & \sim & -0.67 \pm 0.92 & \hspace{2cm} & X & \sim & 0.1 \pm 0.58 
\end{array} \eeq

Explicit calculation \cite{VWX} in ETC models gives the result $X \sim 0$ in all
scenarios. The parameter $V$ is only non-zero when a techni-fermion's mass is of order 
$M_Z$ ($M \sim 50GeV$ $V \sim -0.15N_{TC}$; $M \sim 100GeV$ $V \sim -0.02N_{TC}$), where
$N_{TC}$ is the number of technicolours. 
If V and X both fall to zero the global fit to data for S 
and T is much more restrictive; the  one standard deviation bounds are \cite{Burgess2}

\beq \begin{array}{ccccccc}
S & \sim & -0.5 \pm 0.6 & \hspace{2cm} & T & \sim & -0.3 \pm 0.6 \\
\end{array} \eeq

Calculating V,W and X for a strongly interacting doublet is difficult since these
parameters measure the deviations from a Taylor expansion of the gauge boson
self energies. Chiral models of strong interactions \cite{ChiL} in which the low energy 
effective theory is given as a derivative expansion are, therefore, completely
inadequate. It is reasonable to assume that the strongly interacting results show
the same behaviour as the weakly interacting results.
In our model we shall assume that
the techni-neutrino mass is $\sim 50-100GeV$ so that V is non-zero and the less stringent 
bounds on S and T apply. 

The contribution to the T parameter from a weakly interacting fermion doublet (U,D) with 
momentum independent  mass is given approximately  by the form \cite{Peskin}

\beq T \simeq \frac{1}{12\pi s^2_{\theta_W} c^2_{\theta_W}} \left[ \frac{ (\Delta
m)^2}{M_Z^2} \right] \eeq

\noindent where $s_{\theta_W}$ and $c_{\theta_W}$ are the sine and cosine of the weak
mixing angle and $\Delta m$ the mass splitting within the doublet. We conclude that
techni-fermions with masses much greater than $M_Z$ must be mass degenerate or else
give too large a contribution to the T parameter. For example a doublet with mass splitting
of $150GeV$ contributes $T = 0.41N_{TC}$, of $100GeV$ contributes $T = 0.18 N_{TC}$, 
and of $50GeV$ contributes $T = 0.04 N_{TC}$. 
The T parameter contribution from a strongly interacting doublet can be estimated in 
Dynamical Perturbation Theory \cite{DPT} as

\beq  T = \frac{1.37}{F_{\pi^3}^2} ( F_{\pi^{\pm}}^2 - F_{\pi_3}^2 ) \eeq

\noindent where $F_{\pi^{\pm}}$ and $F_{\pi^3}$ are the charged and neutral techni-pion decay
constants given by \cite{Holdom}

\beq F_{\pi^3} = \frac{N_{TC}}{32\pi^2} \int^{\Lambda^2}_0 dk^2 k^2 \left( \frac{ \Sigma_U^2
                            - k^2 ( \Sigma_U^2)' /4}{(k^2+\Sigma_U^2)^2}  +  U \leftrightarrow D \right) \eeq

\beq F_{\pi^{\pm}} = \frac{N_{TC}}{32\pi^2} \int^{\Lambda^2}_0 dk^2 k^2 \frac{F(k^2) }
                              {(k^2+\Sigma_U^2)(k^2+\Sigma_D^2)} \eeq

\beq \begin{array}{ccc} F(k^2) & = &  (\Sigma_U^2+\Sigma_D^2) - \frac{1}{4} k^2
                 (\Sigma_U^2+\Sigma_D^2)' - \frac{1}{8}[(\Sigma_U-\Sigma_D)^2]' \\
&&\\
&&       -\frac{1}{4} (\Sigma_U-\Sigma_D)(\Sigma_U+\Sigma_D)'(\Sigma_U'\Sigma_D-
       \Sigma_U\Sigma_D')\\
&&\\
&&      +[ \frac{1}{2} k^2(\Sigma_U^2-\Sigma_D^2) - \frac{1}{4}k^2(k^2-\Sigma_U \Sigma_D)
      (\Sigma_U-\Sigma_D) \\
&&\\
 &&     \times (\Sigma_U+\Sigma_D)'] \left(\frac{1+(\Sigma_U^2)'}{(k^2+\Sigma_U^2)} - 
\frac{1+(\Sigma_D^2)'}{(k^2+\Sigma_D^2)} \right) \end{array} \eeq 

\noindent where $\Sigma_U$ and $\Sigma_D$ are the self energies of the fermions and
the prime indicates the derivative with respect to $k^2$. These equations show the same
behaviour as Eqn(2.3) with some enhancement \cite{Holdom} for a given mass splitting. We shall make 
use of them in our analysis of T below.

The contribution to the S parameter from a weakly interacting fermion doublet with 
momentum independent Dirac masses is given  by the form \cite{Peskin}

\beq S_{weak} = \frac{1}{6\pi} \left[ 1 - Y_L \ln \left( \frac{m_U^2}{m^2_D} \right) \right] \eeq

\noindent where $Y_L$ is the left handed doublet's hypercharge. There has been much 
discussion in the literature \cite{NPS} of how this result is affected by the inclusion of strong
interactions for the doublet. In the Non Local Chiral Model (NLCM) of strong interactions in
Ref\cite{NLCM} the S parameter may be expressed as an integral equation over the
techni-fermions' self energies. As these self energies deviate from being momentum 
independent (as is suggested by gap equation solutions \cite{gap}) the contribution 
to S rises in the custodial SU(2) limit. Walking technicolour theories \cite{Walk} and
models with strong ETC interactions (such as we shall have below) which enhance the
high momentum tail of the self energies \cite{gap} will presumably give contributions 
to S that lie between the highest estimate of the NLCM and the perturbative result.
It is also unclear whether custodial SU(2) violation in the high momentum tails of
the doublet's self energy is sufficient to give negative contributions to S.
We shall adopt as an upper bound on the contribution for a
techni-doublet

\beq S_{strong} = N_{TC} [ S_{weak} + 0.05 ] \eeq

\noindent where $N_{TC}$ is the number of technicolours and where we calculate $S_{weak}$ 
using the techni-fermions mass (given by $\Sigma(m) = m$). This result agrees with the
observed data for QCD (the custodial SU(2) limit) and with an analysis of the 
contribution to S from a techni-lepton
doublet with a small Majorana mass perturbing the custodial isospin limit \cite{Maj}. In addition we note
that this result is the most conservative estimate of S in the literature away from the
 custodial isospin limit (it reduces
the negative contributions from doublets with mass splittings). In a one family 
technicolour model such as we shall be considering below in which the techni-fermions
are all mass degenerate we obtain $S = 0.4N_{TC}$ which is in excess of the experimental
limit for all but the most minimal technicolour groups. To reduce this value  we 
require doublets with mass splittings, however, we must be careful not to violate the T 
parameter bound.

These results lead to a one family technicolour techni-fermion mass spectra of the form
\cite{Terning,Maj}

\beq m_Q \sim {\rm degenerate}, \hspace{1cm} m_E \sim 150-250GeV, 
\hspace{1cm} m_N \sim 50-100GeV \eeq

\noindent Perturbatively these doublets would give $S \sim 0.09N_{TC}$, $T \sim 0.3N_{TC}$
and $V \sim -(0.15- 0.02)N_{TC}$. Our non-perturbative upper bound
on S is thus $0.29N_{TC}$. We conclude that this techni-fermion spectrum probably lies
within the experimental constraints for $N_{TC} < 6$. In 
addition we note that a large Majorana neutrino mass for the techni-neutrino gives a
somewhat larger negative contribution to S and a smaller contribution to T \cite{Maj}. 

\subsection{Non-Oblique Corrections}

The ETC gauge bosons responsible for the light fermion masses give rise 
to non-oblique corrections to fermion anti-fermion production rates at 
LEP \cite{ZBB}. If the ETC interactions are orthogonal to the standard model gauge group
then these non-oblique effects serve to correct the left handed fermion 
couplings by

\beq \delta g_L^{ETC}  \sim - \frac{1}{2} \frac{g_{ETC}^2}{M_{ETC}^2} 
F_{\pi}^2  \frac{e}{s_{\theta_W} c_{\theta_W}} I_3 \eeq

\noindent where $g_{ETC}$ and $M_{ETC}$ are the ETC gauge boson 
coupling and mass respectively, $I_3$ is the external fermion's weak
isospin and $F_{\pi}$ is the electroweak symmetry breaking scale.
Only the coupling of the ETC gauge boson, $g^2_{ETC}/M^2_{ETC}$, that is
responsible for the top quark's mass  is  sufficiently large for the experimental 
data to constrain. These non-oblique effects are potentially visible in the 
$Z \rightarrow b{\bar b}$ vertex, measured by the ratio of Z boson decay widths to
$b \bar{b}$ over that to all non$-b \bar{b}$ hadronic final states \cite{ZBB}

\beq \Delta_R = \frac{ \delta ( \Gamma_b / \Gamma_{h \neq b})}{ \Gamma_b / 
\Gamma_{h \neq b}} \sim  \frac{2 \delta g_L g_L}{g_L^2 + g_R^2} \eeq

\noindent where $g_L = \frac{e}{s_{\theta}c_{\theta}} ( - \frac{1}{2} + 
\frac{1}{3}s_{\theta}^2)$, $g_R = \frac{e}{s_{\theta}c_{\theta}} (\frac{1}{3}
s_{\theta}^2)$. 

If the top quark mass ($m_t > 130GeV$) is generated by a perturbative ETC
gauge boson (ie $g^2_{ETC} \sim 1$) then the ETC breaking scale must be 
of order 1TeV. The ETC contribution to $\Delta_R \sim 4\%$ \cite{ZBB,ZBB2}
is approximately
double the maximum experimentally consistent value \cite{Burgess3}. However, if the ETC
coupling is allowed to rise to $40-80\%$ of it's critical coupling 
($g_C^2 = 8\pi^2$)at a 
breaking scale of 10TeV then a physical top mass can be obtained for a 
realistic value of $\Delta_R$ \cite{ZBB3}. We shall, therefore, take the lightest ETC
gauge boson to have mass $M_{ETC} \sim 10TeV$. 

\subsection{Other Experimental Constraints}

There are two additional constraints on ETC models, flavour changing neutral
currents (FCNC), and the large, potentially light, pseudo Goldstone boson spectrum
associated with the  $SU(8)_L \otimes SU(8)_R \rightarrow SU(8)_V$ global 
chiral symmetry breaking of the techni-fermions. We shall breifly review these problems in this section

FCNCs \cite{TC,ETC} arise in ETC models through the interactions of the massive gauge bosons
associated with the breaking $SU(N+3)_{ETC} \rightarrow SU(N)_{TC} +$ three 
light families. Each of the light fermions has an associated ETC coupling,
$g^2_{ETC}/M^2_{ETC}$, given by

\beq g^2_{ETC}/M^2_{ETC} \sim m_f / \Lambda_{TC}^3 \eeq

\noindent where $m_f$ is the fermion's mass. An analysis of the contributions to
FCNCs in Ref\cite{FCNC} assuming that any FCNC involving a particular light fermion
have a coupling at least as small as the calculated value in Eqn(2.13) reveals no
constraints on the model from FCNCs. In addition we note that in models such as
those we discuss below with strong ETC interactions the ETC coupling in Eqn(2.13) 
can be a considerable over estimate and hence FCNCs will be suppressed further.
Thus although the contributions to FCNCs are model dependent models \cite{ETC2} do exist in
the literature which naturally avoid FCNC constraints.

ETC models with a full techni-family give rise to 60 light pseudo Goldstone bosons (PGB)
and 3 massless Goldstone bosons associated with the 63 broken generators of the 
techni-fermions' approximate global chiral symmetry \cite{PGB1}. The 60 PGBs
acquire masses through the standard model and ETC interactions that perturb the
global symmetry group. Calculation \cite{PGB1} of the PGB's masses from the $SU(3)_C \otimes
SU(2)_L \otimes U(1)_Y$ interactions of the techni-fermions suggests that as many 
as 7 PGBs may have masses below the current experimental search limits. However,
the major source of global symmetry breaking in the standard model comes from
the fermion masses generated in ETC models by the ETC interactions. Calculation \cite{PGB2} of the 
contribution to the PGB masses from ETC interactions sufficiently strong to 
generate the observed light fermion mass spectra reveal that all the PGBs will
have masses in excess of the current direct search limits. The only exception is
the neutral PGB with constituent techni-neutrinos. However, neutrino mass
generation is extremely model dependent and in the absence of a convincing
model of the neutrino sector we argue that it is not possible to place an upper constraint
on the PGB mass. These calculations suggest that ETC models are unconstrained 
by the PGB spectrum.

In addition to the usual PGB spectrum the authors of Ref\cite{light} have argued that
when ETC interactions grow close to their critical values there will be additional 
light (relative to $M_{ETC}$), scalar, ETC bound states of the light fermions. These
bound states' masses will fall to $\sim 2m_f$, where $m_f$ is the mass of the constituent
fermion, as the ETC interactions grow to their critical values. The strongest ETC
interactions in our models below ($\sim 80-90\%$ of $g_C$), which will presumably 
give rise to the lightest scalar spectrum, are associated with the top mass generation.
We, therefore, expect the lightest such scalar to have a mass $> 100GeV$.

\section{A Generalized ETC Model}

We wish to study the viability of a range of ETC models without restricting
to any particular scenario. Since we have argued that the experimental
constraints restrict models to an ETC breaking scale of 10TeV or greater it
will be a good approximation to model the ETC interactions by simple four
Fermi operators (we expect higher dimensional operators to be sufficiently 
suppressed). Thus our general model will consist of an SU(N) technicolour
group and, in principle, any number of gauge invariant four Fermi operators
acting on a full techni-family (N,E,$U^c$,$D^c$: c is a colour index). 
In addition we consider the third family 
of fermions which are technicolour singlets but interact with the techni-fermions
by ETC interactions again modelled by four Fermi operators. The technicolour
group becomes strongly interacting at the scale $\Lambda_{TC} \sim 1TeV$
forming techni-fermion condensates and breaking electroweak symmetry.
We shall allow the ETC charges to vary
over all possible values and search for a general model(s) compatible with
the experimental data discussed in Section 3 and the third family 
fermion masses. These solutions will hopefully provide a general basis from
which to build more specific (renormalizable) models.

The ETC interactions in our model can be split in to two catagories, 
sideways and horizontal. Sideways interactions feed the techni-fermion 
condensates down to the light three families of fermions. There are four such
operators connecting the techni-fermions and third family

\beq 
\frac{g_{\nu_3}^2}{M^2_{ETC}} {\bar \Psi}_L N_R \bar{\nu}_{\tau R} \psi_L \hspace{0.5cm}
\frac{g_{\tau}^2}{M_{ETC}^2} {\bar \Psi}_L E_R \bar{\tau}_{R} \psi_L \hspace{0.5cm} 
\frac{g_{t}^2}{M^2_{ETC}} {\bar Q}_L U_R \bar{t}_{R} q_L \hspace{0.5cm} 
\frac{g_{b}^2}{M^2_{ETC}} {\bar Q}_L D_R \bar{b}_{R} q_L \eeq

\noindent where $\Psi = (N,E)$, $\psi = (\nu_{\tau},\tau)$, $Q=(U,D)$ and
$q=(t,b)$. For readers who wish to have a renormalizable ETC model in mind 
these correspond to operators generated by breaking $SU(N+1)_{ETC} \rightarrow
SU(N)_{TC} +$ third family at the scale $M_{ETC} \sim 10TeV$. 

Horizontal interactions correspond to techni-fermion and light fermion self
interactions of the form

\beq \frac{g_f^2}{M^2_{ETC}} {\bar F}_L f_R \bar{f}_{R} F_L \eeq

\noindent where F is the left handed doublet containing the general fermion
f and where there may in general be such an interaction for each fermion in
the model. We might expect the third family fermions and their respective 
techni-fermion counter parts to share quantum numbers and hence horizontal
interactions. Our models will respect
this constraint except when direct top condensation is investigated. 
Again the reader may envision that these interactions are generated 
at the scale $M_{ETC}$ perhaps most naively by the breaking of an additional
U(1) gauge group (allowing for the different fermions within a family 
to have different 
horizontal charges). We also note that all the four Fermi operators will 
have charges below their critical couplings hence we may skip any discussion
of the strong properties of isolated U(1) gauge interactions.

A realistic ETC scenario must agree with experimental data for $m_{\nu_{\tau}}$,
$m_{\tau}$, $m_t$, $m_b$, V,T and S. The general model has 8 independent four
Fermi charges (12 if we allow the third family horizontal interactions
to differ from their techi counter parts) and, therefore, we might expect 
solutions to exist compatible with
the data. In Section 5 we will verify that such a solution exists, however, it is
clearly unpredictive. It is interesting to propose the minimal model in
principle capable of reproducing the fermion mass spectra and test it for mass 
predictions. 

To simplify
the initial analysis of this paper we shall neglect the discussion of the neutrino masses in
the model since their masses do not fit any obvious pattern in relation to the 
other light fermion masses. We effectively assume that there are no right 
handed neutrinos though we maintain right handed techni-neutrinos. The precise
mechanism for suppressing neutrino masses is extremely model dependent. 
In addition since there is no reliable method of estimating the contribution
to the S parameter from strongly interacting doublets we shall simply set the 
techni-neutrino mass to 50-100GeV in the future discussion and assume that a
realistic S parameter is obtained provided the techni-electron mass lies between
150-250GeV as discussed in Section 2. The W and Z gauge 
boson masses will be dominated by the heavier techni-quarks and hence 
neglecting the details of the neutrino sector will have little effect on the 
technicolour dynamics. The T parameter, however, will presumably be dominated 
by the techni-lepton sector as in the techni-fermion spectra discussed in Section 
2.  We assume that the techni-lepton sector contributes $T \leq 1$ and
hence the T parameter contribution from the techni-quarks must at most be a
few tenths.

In addition we note that the CKM matrix elements only significantly vary
from the identity for the first (lightest) family of fermions whose masses
are generated by the weakest ETC operators. We conclude that quark mixings
and CP violation are generated by those weak interactions and, therefore, in
discussion of the heavier two generations of fermions we may neglect the
CKM matrix elements. There is no clear understanding of the 
origin of the CKM matrix elements and hence we wish to neglect their generation 
in this discussion since we wish to make model independent predictions.
Making this approximation will clearly upset any predictions of the 
first family masses which are associated with large mixings and indeed in 
Section 6 we shall see this manifest.

Now we may consider the minimal number of ETC interactions neccessary to
generate the light fermion masses \cite{Mass}

\beq \begin{array}{ccc} m_t = 160 \pm 30 GeV& m_b = 5.0 \pm 0.3 GeV& m_{\tau} 
                              = 1.784 GeV\\
&&\\
m_c = 1.5 \pm 0.2 GeV & m_s = 0.2 \pm 0.1 GeV & m_{\mu} = 0.105 GeV \\
&&\\
m_u =  5 \pm 3 MeV      & m_d = 10 \pm 5 MeV & m_e = 0.51 MeV \end{array} \eeq

\noindent The EWSB scale is set by the technicolour dynamics corresponding to
the scale $\Lambda_{TC}$ at which the technicolour group becomes strongly
interacting. The third family masses are suppressed relative to this scale by a
factor of $\sim 10$, the second family by a further factor of $\sim 10-100$ and
the first family by yet a further factor of $\sim 10-100$. It is natural to associate
each generation with a separate sideways interaction (introducing a single 
additional interaction parameter for each family). The quarks in each family are
more massive than the leptons so we must break the symmetry between them by 
the addition of at least one extra interaction; we shall introduce a single horizontal
interaction for the quarks. Finally we notice that in the heaviest two families 
the top type quarks are more massive than the bottom type
(for the moment we ignore the up down mass inversion since it is associated
with the scale at which the approximation that the CKM matrix is the 
identity breaks down) and hence there must
be an additional interaction on these quarks to break the symmetry between them;
we introduce a single additional horizontal interaction for top type quarks.

There must be a minimum of 5 new interactions in our
model to break the global symmetries  that would otherwise leave the light 
fermions degenerate. Indeed it is hard to imagine how any model of the light
fermion masses could have fewer free parameters than this.

\section{The Devil We Know - The Gap Equation}

Before we can discuss the success or failure of scenarios such as those discussed
in Section 3 we must have a reliable method of calculating physical quantities in
strongly interacting theories. The infinite tower of Schwinger Dyson equations
are untractable so it is traditional to truncate the tower after the fermion two 
point function and replace other propagators and vertices with the 
perturbative Feynman rule. We then obtain the two gap equations \cite{gap} for the fermion
self energy from SU(N)
gauge interactions (in Landau gauge and with a running gauge coupling)
 and four Fermi interactions respectively \vspace{-0.5cm}

\beq \Sigma(p)  =  \frac{3 C(R)}{4 \pi} \int^{\Lambda^2}_0 \alpha(Max(k^2,p^2)) 
\frac{k^2 dk^2}{Max(k^2,p^2)} \frac{\Sigma(k)}{k^2+\Sigma^2(k)} \eeq

\beq \Sigma(p) = \frac{g^2}{8\pi^2\Lambda^2} \int^{\Lambda^2}_0 k^2 dk^2 
\frac{\Sigma(k)}{k^2+\Sigma^2(k)}  \eeq

$\left. \right.$

\noindent where C(R) is the casimir operator of the fermion's representation of 
the  gauge group, $\alpha$ the running gauge coupling, g the four Fermi
interaction strength, and $\Lambda$ the UV cut off.

The major success of these gap equations is that they show chiral symmetry breaking
behaviour \cite{gap}. In each case there is some critical coupling below which the solution
to the equation is $\Sigma(k^2) = 0$ and above which $\Sigma(k^2) \neq 0$. 
Clearly, however, in the case of the gauge coupling the precise value of the critical
coupling and the form of the solution depend upon the form of the running 
of the coupling both in the high momentum regime (where the  \newpage \noindent running may be
calculated in perturbation theory and is known to depend on the number of
interacting fermions) and in the non-perturbative regime. 

In order to investigate the consistency of solutions within the gap equation 
approximation let us consider the minimal predictive ETC model proposed in
Section 3. The gap equations for the techni-family and third family are

\beq \begin{array}{ccl}
&&\end{array} \eeq

\newpage

\noindent where D(R) is the dimension of the techni-fermions' representation
under the technicolour group. 

The scale $\Lambda_{TC}$ is determined by 
requiring the correct Z mass which is given by the techni-pion decay constant, $F_{\pi}$,
in
Eqn(2.5-7). For simplicity we neglect the mass splitting within the lepton doublet in the
calculation of $F_{\pi}$; since the Z mass is dominated by the techni-quark
contribution to $F_{\pi}$ this will introduce only small errors and allows us to
avoid the complication of specifying the neutrino sector. The three four Fermi
couplings are determined, for a given value of $M_{ETC}$, by requiring 
that the correct tau, top and bottom masses are obtained as solutions. We tune to 
two significant figures in the fermion masses and use $m_t \sim 170 GeV$ as a representative value.
We cut the integrals off at $M_{ETC}$.

The dependence of the solutions on the $N_{TC}$, $M_{ETC}$ and
the running of the coupling both in the perturbative and non-perturbative 
regimes may now be investigated. We begin by allowing the technicolour
coupling to run according to the one loop $\beta-$function result above
$\Lambda_{TC}$ and cut off the running below $\Lambda_{TC}$ 

\beq \begin{array}{cccl}
                   \alpha(q^2) & = & 2 \alpha_C     &  \hspace{0.5cm} q^2< \Lambda_{TC} \\
                   &&\\
                   \alpha(q^2) & = & \frac{2\alpha_C}{1+2\alpha_C\beta \ln(q/\Lambda_{TC})}
                   & \hspace{0.5cm}  q^2> \Lambda_{TC} 
\end{array} \eeq

\noindent where $\alpha_C$ is the critical coupling in the fixed point theory 
($\alpha_C = \pi/3C(R)$). We set $\beta=1$, a typical running value and 
$M_{ETC} = 10 TeV$. We assume that the techni-fermions lie in the
fundamental representation of the technicolour group. In Fig 1
we show results for the techni-fermion self energies as a function of momenta
for $N_{TC} = 3$ and 6. In Fig 2  we display the dependence of the
techni-up quark's self energy in the $SU(3)_{TC}$ scenario to changes in the 
$\beta-$function for $M_{ETC} = 10TeV$. If the $\beta-$function falls
below 0.22 then $\Lambda_{TC}$ must be reduced below 100GeV which is
presumably unphysical. In Fig 3
we show the low energy structure of the techni-up quark's self energy for 
varying ETC scales ($M_{ETC}
= 5,10$ and $50TeV$) again with $N_{TC} = 3$ and $\beta = 1$.
The couplings that satisfy all these solutions are given in Table 1. \vspace{3in}
\newpage
\begin{center} 

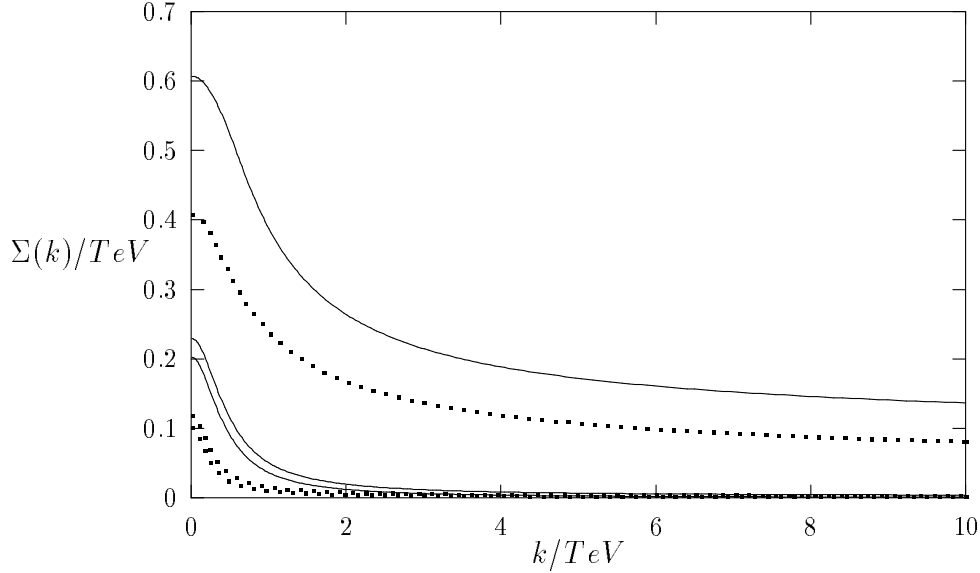
Fig 1: The solutions to the gap equations of Eqn(4.3) for the techni-fermion
self energies with $N_{TC} =3$ (solid curves) and $N_{TC} =6$ (dashed curves).
$M_{ETC} = 10TeV$ and $\beta = 1$.
In each case the highest curve is the techni-up self energy, the middle 
curve the techni-down self energy and the lower curve the techni-electron self
energy. The solutions are given by tuning the couplings to $M_Z,m_t,m_b$ and 
$m_{\tau}$.\vspace{3in} \end{center}

\begin{center} 

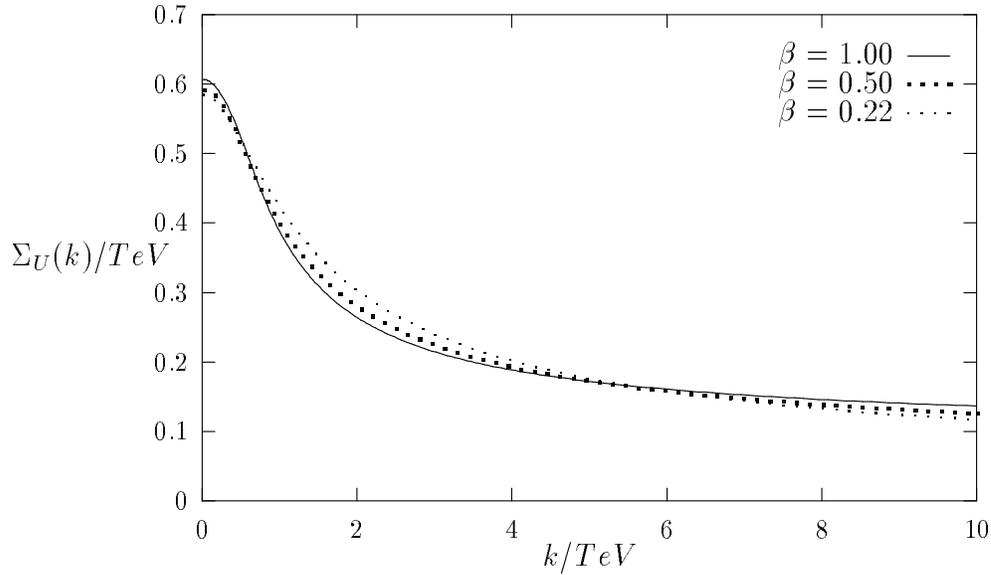
Fig 2: Dependence of gap equation solutions in Eqn(4.3) for the techni-up self
energy on the
technicolour $\beta-$function with $N_{TC} = 3$ and $M_{ETC} =10TeV$. \end{center}
\newpage
\begin{center} 

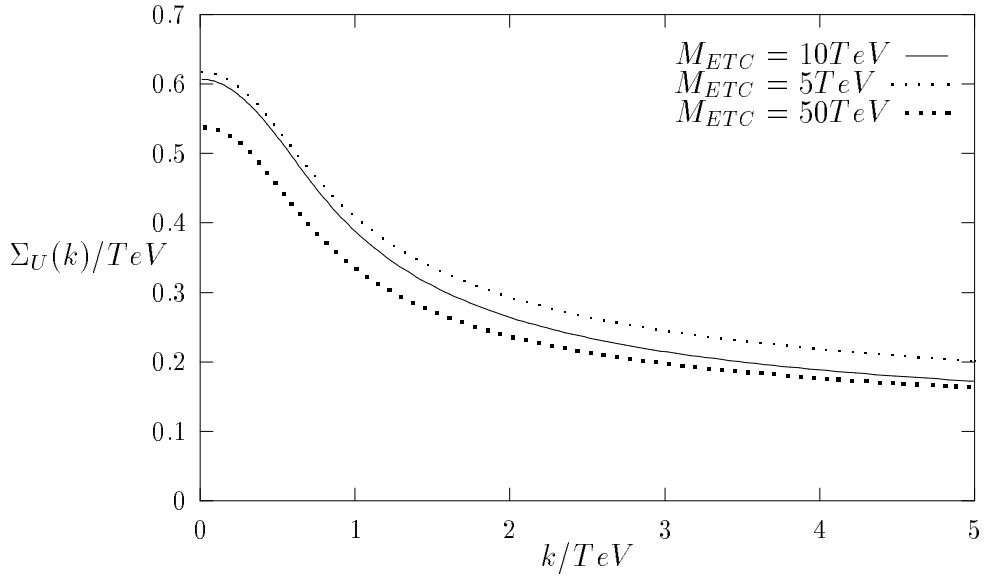
Fig 3: Dependence of gap equation solutions for the techni-up self energy 
in Eqn(4.3) on $M_{ETC}$ 
 with $N_{TC} = 3$ and  $\beta = 1$. \end{center}

The ansatz for the running of $\alpha$ in the non-perturbative regime in Eqn(4.4)
is only determined in as much as it must be finite at $q = 0$. In Fig 4 we compare the 
effects of two extreme choices for this regime. The first ansatz assumes that the 
coupling flattens out quickly at low momenta having the form of Eqn(4.4) but taking
a maximum value of $1.5\alpha_C$. This ansatz is probably an underestimate of
the coupling strength since there is a large discrepancy between $\Lambda_{TC}$
and $\Sigma(0)$. The second ansatz assumes that outside the perturbative regime
the coupling rises sharply from $\alpha_C$ to a maximium value of $3\alpha_C$

\beq \begin{array}{cccl}
                   \alpha(q^2) & = &  3\alpha_C     & \hspace{0.5cm}  q^2< \Lambda_{TC} \\
                   &&\\
                   \alpha(q^2) & = & \frac{\alpha_C}{1+\alpha_C\beta \ln(q/\Lambda_{TC})}& 
                   \hspace{0.5cm}  q^2> \Lambda_{TC} 
\end{array} \eeq

\noindent this presumably is a somewhat over estimate of the coupling strength.\newpage

\begin{center} 

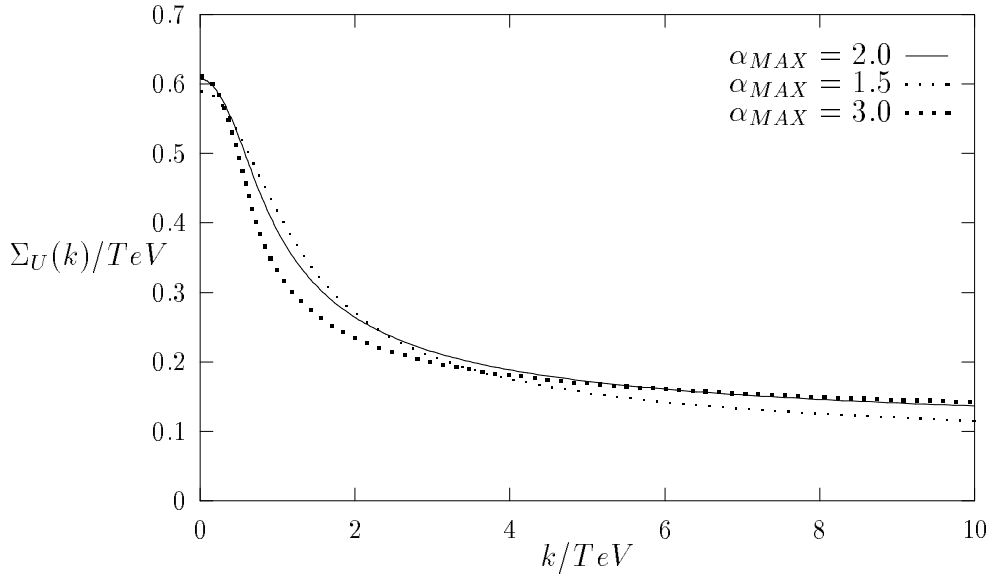
Fig 4: Dependence of gap equation solutions on the non perturbative running. Details 
of the coupling ansatzs are given in the text. $N_{TC} =3$, $\beta = 1$and $M_{ETC}=10TeV$.
 \end{center}

We observe that 
in each case the self energy solutions have the same general form though it is
clearly impossible to distinguish the solutions phenomenologically. Although there
is some variation in the shape of $\Sigma(k^2)$ the area under $\Sigma(k^2)$ 
that contributes to the light fermion masses are fixed (by the requirement that they
give the correct Z mass) at least up to errors of at most order one. We therefore
expect the light fermion masses we calculate in the gap equation approximation to
a given ETC model to at least be representative of the rough pattern of masses the
theory would produce. However, the precision electroweak parameters are 
plagued by error in this approximation. The T parameter is a measure of one
percent differences between our calculated values of $F_{\pi^3}$ and $F_{\pi^{\pm}}$
which correspond to integrals over the self energies. Clearly this level of precision 
is not provided for. The calculated values for the techni-quark contribution to T
in each of the above scenarios  is given
in Table 1 and vary between T=8.9 and T=24.2! Similarly we have argued that to 
achieve a realistic value of S and 
V we require the techni-electron mass (determined by the condition 
$\Sigma(M_E) =M_E$) lies in the range $150 - 250 GeV$. The calculated value of $M_E$
is given in Table 1 also and again we see a large variation, $M_E = 90-260GeV$. 
We shall only \newpage \noindent be able to argue about the gross features of the techni-fermion
spectra and on where these are compatible with the realistic mass spectra in Eqn(2.10).

\begin{center}
\begin{tabular}{|c|cccc|cccc|cc|}
\hline
 & &&&& &&&& &\\
 & $N_{TC}$ & $\alpha_{MAX}$ & $\beta$ & $M_{ETC}$ & $\Lambda_{TC}$ 
                             & $g_3/g_C\%$ & $g_Q/g_C\%$ & $g_t/g_C\%$ & $T_Q$ & $M_E/GeV$ \\
 & &&&& &&&& & \\ \hline
 & &&&& &&&& &\\
 & 3 & 2 & 1.00 & 10 & 0.60 & 40.0 & 51.8 & 52.66 & 16.4 & 170 \\
 & &&&& &&&& & \\ \hline
 & &&&& &&&& &\\
FIG 1 & 6 & 2 & 1.00 & 10 & 0.20 & 13.4 & 33.2 & 35.1 & 20.4 & 90 \\
 & &&&& &&&& & \\ \hline
 & &&&& &&&& &\\
 & 3 & 2 & 0.75 & 10 & 0.50 & 41.6 & 49.5 & 50.1 & 17.5 & 160 \\
 & &&&& &&&& &\\
FIG 2 & 3 & 2 & 0.50 & 10 & 0.35 & 45.0 & 44.5 & 46.1 & 19.5 & 140 \\
 & &&&& &&&& &\\
 & 3 & 2 & 0.22 & 10 & 0.10 & 52.9 & 29.3 & 36.0 & 24.2 & 90 \\
 & &&&& &&&& & \\ \hline
 & &&&& &&&& &\\
FIG 3 & 3 & 2 & 1.00 & 5 & 0.52 & 30.4 & 50.64 & 60.8 & 19.0 & 150 \\
 & &&&& &&&& &\\
 & 3 & 2 & 1.00 & 50 & 0.49 & 70.7 & 17.8 & 14.5 & 15.8 & 140 \\
 & &&&& &&&& & \\ \hline
 & &&&& &&&& &\\
FIG 4 & 3 & 3 & 1.00 & 10 & 0.60 & 36.0 & 55.1 & 57.7 & 8.9 & 260 \\
 & &&&& &&&& &\\
 & 3 & 1.5 & 1.00 & 10 & 1.10 & 48.4 & 38.3 & 47.8 & 22.9 & 100 \\
& &&&& &&&& & \\ \hline  \end{tabular}

Table 1: Numerical values of the couplings and scales used to plot Fig 1-4.
The non-perturbative  ansatz for the technicolour coupling is indicated by
the maximum value $\alpha_{MAX}$. The four Fermi couplings are given as 
percentages of the critical coupling ($g_C^2 = 8\pi^2$). $T_Q$ is the contribution
to T from the techni-quarks. Solutions are obtained by tuning parameters to
give the correct Z mass, $m_{\tau}$, $m_b$ and $m_t$.

\end{center}

Finally we note that even if the gap equations are not a realistic approximation to
the underlying Schwinger Dyson equations they still provide a parameterization 
of the techni-fermions' self energies. Thus whilst the gap equation couplings may not be
physical the existence of gap eqaution solutions consistent with the experimental 
data is indicative that couplings exist in the full theory also compatible with the
data.

\section{Successful Scenarios}

Our analysis in Section 4 of the minimal predictive model proposed in Section 3
suggests that the techni-quarks in such a scenario give rise to too large a
contribution to the T parameter ($T_Q \sim 15$, see Table 1). It is interesting to note
however that the techni-fermion self energies (in Fig 1) show the general pattern
of the realistic mass pattern in Eqn(2.10) except for the overly large splitting between
the techni-up and techni-down quarks. In this section we present two scenarios
in which the techni-up techni-down mass splitting lies within experimental 
constraints, one model is completely unpredictive the other is a variation on the minimal
predictive model with direct top condensation.

\subsection{An Existence Proof}

In principle the ETC couplings in the generalized ETC model described in Section 3
need not be related and we obtain the gap equations \newpage

\beq \begin{array}{ccl}
&&\end{array} \eeq

The top and bottom quark masses within this general model are determined by 
their separate sideways interactions. Although the top and bottom masses 
feed back into the techni-fermions' self energies tending to enhance the techni-up
self energy it is clear that the separate  horizontal interactions on the top and 
bottom type quarks can be used to enhance the techni-bottom self energy to
oppose this custodial SU(2) violating effect. We can tune a set of couplings 
to give $T_Q=0$ and which correctly describe the Z, tau, top and bottom  \newpage masses 
eg a scenario with $g_E = g_U = 0$:

\begin{center}
\begin{tabular}{|cccccccccc|}
\hline
  &&&& &&&& &\\
$N_{TC}$ & $\alpha_{MAX}$ & $\beta$ & $M_{ETC}$ & $\Lambda_{TC}$ 
        & $g_{\tau}/g_C\%$ & $g_b/g_C\%$ & $g_t/g_C\%$ &$g_D/g_C\%$ &$T_Q$  \\
 &&&& &&&& & \\ \hline 
 &&&& &&&& & \\
 3 & 2.0 & 1.00 & 10 & 0.5 & 48.4  & 5.9 &  70.1 & 85.5  & 0.0 \\
 &&&& &&&& & \\ \hline 
\end{tabular}
\end{center} 

\noindent which give the techni-fermion masses 

\beq M_U \sim 400GeV, \hspace{1cm} M_D \sim 400GeV,\hspace{1cm} M_E \sim 140GeV \eeq

\noindent Such a scenario is consistent with the techni-fermion mass spectrum
in Eqn(2.10) and hence with all available 
experimental data. The  renormalizable models of Ref\cite{ETC2}
can  give rise to precisely this spectrum of ETC interactions, however, the degeneracy of the 
techni-quarks (and hence the low T parameter) arises from a conspiracy in the
four Fermi couplings which seems unnatural. Nevertheless this scenario does provide 
an existence proof for ETC models.

\subsection{Direct Top Condensation}

The minimal predictive model of Eqn(4.3) fails because the techni-up self energy
must be enhanced by too much relative to the techni-down in order to generate the
top bottom mass splitting. Recently there has been much discussion in the 
literature of direct top condensation \cite{TMSM} giving rise to the large top mass. Whilst
top condensation on its own is plagued by difficulties of fine tuning in order not
to generate too large a top mass (ruled out by the T parameter measurements) when
the top is not the major source of EWS breaking such fine tuning problems need
not exist. We can construct an ETC model with top condensation simply by 
removing the horizontal interaction on the techni-up quark in the minimal 
predicitive model. Since the large top mass is no longer generated by the
sideways ETC interactions there is less constraint upon the ETC breaking scale, $M_{ETC}$,
from the $Z \rightarrow b {\bar b}$ vertex measurements. We shall allow
$M_{ETC}$ to fall to 5 TeV. The gap equations are then

\beq \begin{array}{ccl}
& & \end{array} \eeq

In Table 2 we show some solutions to these equations and their predictions for the
contribution to the T parameter from the techni-quarks. \newpage

\begin{center}
\begin{tabular}{|cccc|cccc|cc|}
\hline
 &&&& &&&& &\\
 $N_{TC}$ & $\alpha_{MAX}$ & $\beta$ & $M_{ETC}$ & $\Lambda_{TC}$ 
                             & $g_3/g_C\%$ & $g_Q/g_C\%$ & $g_t/g_C\%$ & $T_Q$ & $M_E/GeV$ \\
 &&&& &&&& & \\ \hline
 &&&& &&&& &\\
 3 & 2 & 1.00 & 10 & 1.10 & 20.3 & 51.2 & 83.8 & 0.71 & 320 \\
 &&&& &&&& &\\
 3 & 1.5 & 1.00 & 10 & 2.65 & 21.4 & 39.4 & 89.1 & 2.92 & 225 \\
 &&&& &&&& &\\
 3 & 3 & 0.95 & 10 & 0.95 & 20.3 & 58.0 & 79.9 & 0.20 & 410 \\
 &&&& &&&& & \\ \hline 
 &&&& &&&& &\\
 3 & 2 & 0.50 & 10 & 0.78 & 19.7 & 47.3 & 86.1 & 1.00 & 300 \\
 &&&& &&&& &\\
 6 & 2 & 1.00 & 10 & 0.45 & 24.2 & 48.4 & 82.4 & 5.08 & 185 \\
 &&&& &&&& & \\ \hline
 &&&& &&&& &\\
 3 & 2 & 1.00 & 5 & 1.07 & 13.5 & 47.3 & 87.1 & 0.09 & 300 \\
 &&&& &&&& &\\
 4 & 2 & 1.00 & 5 & 0.85 & 12.4 & 47.3 & 87.2 & 0.11 & 270 \\
 &&&& &&&& &\\
 5 & 2 & 1.00 & 5 & 0.65 & 13.5 & 46.1 & 87.8 & 0.24 & 230 \\
 &&&& &&&& & \\ \hline  \end{tabular}
 
Table 2: Numerical values of the couplings and scales of solutions to Eqn(5.3).
The non-perturbative  ansatz for the technicolour coupling is indicated by
the maximum value $\alpha_{MAX}$. The four Fermi couplings are given as 
percentages of the critical coupling ($g_C^2 = 8\pi^2$). $T_Q$ is the contribution
to T from the techni-quarks. Solutions are obtained by tuning parameters to
give the correct Z mass, $m_{\tau}$, $m_b$ and $m_t$.

\end{center}

The solutions with a low ETC scale seem consistent with the techni-fermion 
mass spectrum proposed in Section 2 though the techni-electron mass is 
somewhat high. Within the gap equation approximation it is certainly
not possible to discount this scenario so we shall consider it a successful
ETC model.

\section{Quark Mass Postdictions}

We have argued in Section 3 that a model of EWSB and the third family masses 
(excluding neutrinos) must have at least four couplings and hence can not be
``postdictive" of the third family masses. However, it is conceivable that only one 
additional  parameter need be added to generate the second family masses (a
parameter that suppresses the second family masses relative to the third) since
quark lepton and custodial isospin symmetry breaking already exist in the model.
Similarly one additional parameter might suffice to suppress the first family masses
below the second but of course our neglection of the CKM matrix elements which 
are substantial
for the first family makes this seem less likely to be successful. In this section we 
investigate the possibility of such postdiction in the scenarios we have 
discussed above.

The ``existence proof" scenario does not lend itself to postdiction since to follow
the pattern of the model of the third family masses we could simply introduce 
additional sideways interactions for each new light fermion sufficient to
generate their mass. There are no constraints on the couplings so they are
unpredicitive. The first and second family masses are at least two orders of magnitude
smaller than the techni-fermion masses and hence any feedback of the light
two families masses into the techni-fermion self energies are negligible and do
not upset our calculations of S and T. Although unpredictive the scenario 
still provides an exisistence proof of a realistic ETC model.

The top condensation scenario however is potentially predictive as described above.
We introduce the additional sideways interactions \vspace{-0.5cm}

\begin{eqnarray}
&&\nonumber\end{eqnarray} 
\newpage

\noindent which we would expect to be generated if there was a single breaking 
scale associated with  each of the first and second  families in the breaking of 
$SU(N+3)_{ETC} \rightarrow
SU(N)_{TC} +$ three families. Again the feedback of the first and second family masses
to the techni-fermions and third family are negligible. We set the coupling strength of the 
new sideways interaction by requiring that we generate the correct muon and electron masses.
The up, down, charm and strange quark masses are now predictions of the model. Explicitly

\beq      \begin{array}{ccccccc} 
\Lambda_{TC} &{\rm determined} \hspace{0.1cm} {\rm by} &     M_Z &\hspace{1cm} & g_t &{\rm determined} \hspace{0.1cm} {\rm by} & m_{t}\\
&&\\
g_3 &{\rm determined} \hspace{0.1cm} {\rm by} & m_{\tau} & & g_2 &{\rm determined} \hspace{0.1cm} {\rm by} & m_{\mu}\\
&&\\
g_Q &{\rm determined } \hspace{0.1cm} {\rm by} & m_{b} & & g_1 &{\rm determined} \hspace{0.1cm} {\rm by} & m_{e} \end{array} \eeq

\noindent Although the predictions of the model are clear cut our ability to calculate 
is limited as discussed in Section 4. The gap equation solutions are, however, 
moderately well bounded since the integrals over the techni-fermion's self energies
are fixed to a good degree by the imposed requirements that they correctly give
the Z, tau, bottom and top masses. We shall quote the range of predictions from all 
the coupling values in Table 2 as an estimate of our theoretical errors.  We obtain \newpage

\beq \begin{array}{cc}  m_c = 1.5 \pm 0.8GeV,\hspace{1cm} & m_s = 0.32 \pm 0.02 GeV\\
&\\ 
m_u = 6.6 \pm 3.7 MeV ,\hspace{1cm} & m_d = 1.5 \pm 0.2 MeV \end{array} \eeq

We immediately notice that these predictions are in surprisingly good agreement with
the observed mass spectra except for the down quark. The failure to predict the down quark 
mass however is to be expected since we have neglected the generation of the CKM matrix 
which has large elements for the first family. Conservatively we can conclude that ETC
models with the minimal number of ETC interactions that are sufficient to break the global
symmetry of the light fermions in the observed pattern seem capable of reproducing the pattern of
the observed light fermion mass spectrum.

\section{Conclusions}

The precision data from LEP \cite{Burgess3} has provided tight constraints on the form of models of EWSB.
It has been argued \cite{Terning,Maj} that technicolour models with a single techni-family with a 
light techni-neutrino and degenerate techni-quarks give contributions to the S,T and V
parameters that lie within the experimentally allowed bands. If the top mass is generated
by strong ETC interactions broken above $10TeV$ then the model will lie within
the experimental limits on non-oblique corrections to the $Zb{\bar b}$ vertex as well \cite{ZBB3}. 
As a first step towards a fully renormalizable, predictive model of EWSB we have investigated
whether an ETC model can be compatible both with the precision data and the light fermion
masses. To make this investigation we have used a generalized one family ETC model in
which the ETC interactions are represented by four Fermi interactions.

To calculate within this generalized model we have used the gap equation approximation
to the Schwinger Dyson equations. Unfortunately even within the gap equation
approximation the solutions for the techni-fermions self energies, $\Sigma(k^2)$, are dependent on
the precise form of the running of the technicolour coupling. The technicolour dynamics
are fixed to some degree by the requirement that the model gives rise to the correct
Z boson mass (given by an integral equation over the self-energies). Calculation of
the light fermion masses (also given by integral equations over the self-energies)
are, therefore, moderately stable. However, the precision electroweak variables are very
sensitive to shifts in for example $\Sigma(0)$ and are hence less well determined. 
Nevertheless we have argued that couplings exist in the generalized ETC model that
very plausibly fit the experimental constraints.

Two scenarios in the generalized ETC model have been found consistent with the 
precision data and the third family fermion masses. The first is an unpredictive 
existence proof in which sufficient ETC couplings are included that the fermion 
mass spectra may be tuned to match the data. The second scenario contains
what we have argued is the minimum number of different strength ETC interactions required
to break the global symmetry on the third family in the observed pattern. This
model achieves a sufficiently large top mass by direct top condensation.

In order to obtain a large top mass in these models the ETC interactions must be
tuned close to their critical values. The ``fine tuning" is at worst of order $10\%$,
corresponding in our results to our need to quote ETC couplings to three 
significant figures in order to tune to two significant figures in the light fermion
masses. In fact the tuning is only this severe for the ETC couplings that generate 
the top mass. This degree of tuning may not be unnatural since gauge couplings
naturally run between their critical value, $g_C$, and $\sim 0.1 g_C$ over many 
orders of magnitude of momentum. Clearly any greater degree of fine tuning which,
 for example, would  be associated with significantly increasing $\Lambda_{ETC}$,
would be unsatisfactory.

The top condensing scenario may be minimally extended to the first and second
families. The model then makes predictions for the up, down, charm and strange
quark masses. Our calculation of these masses shows that the charm, strange and up
quark mass predictions are consistent (up to errors due to uncertainty in the gap equation
approximation) with the experimental values. The model does not reproduce the
up down mass inversion observed in nature but we have argued that this is the
result of our neglection of the mechanism for the generation of the CKM matrix 
which has large elements for the first family quarks. In addition we have neglected
a discussion of the neutrino sector since their masses do not fit
any obvious pattern in the fermion mass spectra. In this paper we have concentrated 
on predictions which are potentially generic to ETC models. Clearly it would be of interest
to continue the analysis to models of neutrino masses and the CKM matrix but 
such analysis would only serve to confuse the cleaner model of quarks and
charged leptons.

Hopefully the successes of the generalized ETC model here will be translatable
to a renormalizable ETC model. In this respect the proposal in Ref\cite{gap} that
the quark lepton mass splittings may result from QCD interactions, corresponding
to $g_Q^2 \rightarrow \alpha_{QCD}$ in the top condensate scenario, 
is appealing. At the EWSB scale $\alpha_{QCD}(M_Z^2)/\alpha^{crit}_{QCD} \sim 15\%$.
Our analysis suggests (see Table 2) that $g_Q/g_C$ needs to be of order $50\%$
however. The value of $g_Q/g_C$ can be reduced (see Table 1) by increasing the maximum value 
the technicolour coupling reaches in the non-perturbative regime , or
by increasing $N_{TC}$ or $\Lambda_{ETC}$ or finally by decreasing the
technicolour $\beta-$function towards a walking value. Unfortunately each of
these changes tends to increase the T parameter contribution from the techni-quarks.
The uncertainties in the gap equation analysis though does not preclude the
possibility.

We conclude that our unpredicitve model provides an existence proof that ETC 
models exist which satisfy the stringent precision measurement bounds. The
scenario with direct top condensation provides the tantalizing possibility 
that ETC models can be constructed that are predicitive. \vspace{1in}

\noindent {\Large \bf Acknowledgements}

The author would like to thank the SERC for supporting this work.

\newpage

\end{document}